\documentstyle[12pt]{article}
\def\1{\mbox{I\hspace{-.15em}1}}

\def\b{\begin{equation}}
\def\e{\end{equation}}
\def\bee{\begin{enumerate}}
\def\eee{\end{enumerate}}
\title{Creation of Non-minimal Coupled Particle\\ in Asymptotic de-Sitter Background}
\author{E. Yusofi$^{1}$\thanks{e-mail:E.yusofi@iauamol.ac.ir
} and M. Mohsenzadeh $^{2}$\thanks{e-mail:
mohsenzadeh@qom-iau.ac.ir}}
\begin{document}

\maketitle {\it \centerline{\it $^{1}$ Department of Physics, Ayatollah Amoli Branch, Islamic Azad University, Amol, Mazandaran, Iran}
\centerline{\it P.O.BOX 678, Amol, Mazandaran }  \centerline{$^2$
Department of Physics, Qom Branch, Islamic Azad University, Qom, Iran}}

\begin{abstract}
A general form of quasi-de Sitter(dS) modes in a dynamical background is used to study the creation of particle during the inflation. Actually, by considering the general form of inflaton field equation as a function of the Hankel function index and by using the Planck 2015 constraint on spectral index, we obtain the possible new constraints for the values of coupling constant in the era with quasi-dS space-time. Then, we explicitly calculate the general form of expectation value of the created particles in terms of the Hankel function index and the conformal time. As an important result, we see that the number of created particles and the value of coupling constant can be dependent to the selection of the background space-time and it's dynamics. Our result is general and confirm the conventional special results for the Minkowski and dS background.

\end{abstract}


\section{Introduction and Motivation}
The models of particle creation in the expanding background have received considerable attention in the literature \cite{Bir2, Ful3} and the most important one is the seminal paper of Parker in the late 1960s \cite{Par4, Par5, Par6}. It is well known that the theory of cosmology is governed by the general relativity in curved space-time and if one attempts to build a quantum field theory in curved space-time, general coordinate invariance imposes some ambiguity effects, for example, there is no unique notion for the vacuum state in a generic curved space-time. Immediately, this imposes the ambiguity on the concept of particle since it becomes observer dependent. At every point of space-time, there is a physical prescription which defines an observer frame and each observer may define for him/herself a special vacuum in any frame. Since the vacuum of two observers at two different points of the curved space-time manifold will be different, the positive frequency modes of observer $O_{one}$ or $O_{two}$ can actually be written as a combination of the positive and negative frequency modes of another observer. Therefore, from the point of view of $O_{two}$ , the vacuum state of $O_{one}$ will contain some particles and vice-versa. This is in fact the basic concept about the particle creation that Parker first time offered \cite{Bir2, Mot5}. We are going to use this interesting idea, as a main concept in the section 4.3. of this paper.\\
Recent cosmological observations indicate that the standard inflationary paradigm consisting of a single scalar field undergoing a slow-roll dynamics which could give an excellent description of the early Universe \cite{Obs6}. In particular, the metric perturbations that give rise to the CMB temperature anisotropy and drive the growth of large scale structure are thought to have arisen from quantum fluctuations of scalar field in the vacuum state with the exponential growth of the scale factor stretching the wavelengths of these fluctuations from the micro to the macro scale \cite{Alb7}. The inflaton field is then quantized by choosing a particular mode functions and one may use this selected mode to define the vacuum state.\\
In the standard method, we should choose the linear combinations of mode functions that match with the vacuum mode functions in the short distance limit. This is actually the cornerstone of the Bunch Davies (BD) mode \cite{Bun8} which is selected in accordance with the asymptotic flat space-time \cite{Alb7}. Not only the inflationary universe is not flat but also it has dynamics and it is well known that the early universe can be described with approximate dS space-time \cite{Lin9, Lin10}, thus the study of quantum theory of fields and cosmology in the dS background are well motivated. Noting that our universe at the first approximation can be described by the dS space-time, however, it is certainly conceivable that as we go to shorter distances in early time, one should consider a dynamical Hubbel parameter in quasi-dS background.\\
Since, the recent data from Planck full mission temperature and a first release of polarization data on large angular scales measure the spectral index of curvature perturbations, impose a constraint on the value of scalar spectral index approximately to be  $n_s = {0.968} \pm {0.006}$ at 95\% CL \cite{Obs6} and this observationally constraint motivate us to use initial quasi-dS modes. Therefore, by using of asymptotic expanding of the Hankel function for the selecting of initial quantum modes, we proposed excited-dS modes as the general fundamental modes during inflation\cite{man1, Plb10, Han11, man16}, and by using these alternative non-trivial modes, we obtained the higher order trans-Planckian corrections and scale-dependency of power spectrum \cite{Plb10, man1}. \\
In this work, we show that the departure from dS space-time to quasi-dS space-time can lead to a non-minimal coupling constant between scalar field and gravity in the action. Then, we calculate the number of created non-minimal coupled particles and scale dependency of it resulting from initial quasi-dS modes in comparison with conventional pure dS mode. Actually, In many situations of physical and cosmological interest, non-minimal coupling of the scalar field with scalar curvature is necessary. There are several reasonable grounds to include an explicit non-minimal coupling term in the action\cite{Far12, Far13, Far14, Far15, Far18, Far19}. Also, in the most scenarios of inflation, it turns out that a non-zero value of the coupling constant is unavoidable \cite{Far13}.\\
The rest of this paper proceed as follows: In Sec. 2, we review quantum particle creation in an expanding background. In Sec. 3, motivated from the constraint on scalar spectral index, we propose quasi-dS background instead of dS background by considering the asymptotic expansion of the Hankel functions. In Sec. 4, we provide an observational constraint for the non-zero values of coupling constant in quasi-dS background and we calculate the spectra of created particles in terms of conformal time $\tau$ and Hankel function index $\nu$. We terminate the paper with discussions about particle creation in the some specific background, specially Mikowski, dS and quasi-dS space-time with conformally, minimally and non-minimally coupled inflation at the early universe.
\section{Quantum Particle Creation in Expanding Background}
The general theory of quantum particle creation in expanding background is developed in details in Ref.s \cite{Bir2, Ful3}.
In a curved space-time a real scalar field $\phi$ of mass $m$ is described by the action \cite{Bir2, Ful3, Muk12}
\begin{equation}
 \label{equ8}
S=\frac{1}{2}\int d^4x\sqrt{-g}\Big(g^{\rho\sigma}\partial_{\rho}{\phi}\partial_{\sigma}{\phi}-m^2\phi^2-\xi{R}\phi^2\Big),
\end{equation}
where $R$ is the Ricci scalar and $\xi$ is a constant parameter that characterizes the coupling between the inflaton and the graviton. In terms of the conformal time $ \tau $, the metric tensor $g_{\mu\nu}$ is equivalent to the Minkowski metric $\tau_{\mu\nu}$ conformally, so that the metric given by
\begin{equation}
 \label{equ81}
ds^{2}=a^{2}(\tau)({d\tau}^2-{d\textbf{x}}^2),
\end{equation}
where $a(\tau)$ is the cosmological conformal scale factor. Writing the field $ \phi(\tau,\textbf{x}) = a(\tau)^{-1}u(\tau,\textbf{x})$ , the equation of motion becomes
\begin{equation} \label{equ82}
u''-\nabla^{2}u+(m^{2}a^{2}+(6\xi-1)\frac{a''}{a})u=0,
\end{equation}
and for a massless minimally coupled field, equation (\ref{equ82}) turns to,
\begin{equation} \label{equ9}
u''-\nabla^{2}u-\frac{a''}{a}u=0,
\end{equation}
where the prime denotes derivatives with respect to the conformal time $\tau$.\\
The quantization follows by imposing commutation relations for the field operator
$\hat{u}$ and its momentum canonically conjugate $\hat{\pi}$,
\begin{equation} \label{equ8}
[\hat{u}(\tau,\textbf{x}), \hat{\pi}(\tau,\textbf{y})]=i\delta(\textbf{x}-\textbf{y}).
\end{equation}
The creation and annihilation operators $\hat{a}_\textbf{k}^\dagger$ and $\hat{a}_\textbf{k}$ can be introduced when the field operator $\hat{u}$ is expanded as,
\begin{equation}
\hat{u}=\frac{1}{2}\int\frac{d^3\textbf{k}}{(2\pi)^{3}}\Big(\hat{a}_\textbf{k}u_k(\tau)e^{i\textbf{k}\cdot\textbf{x}}+ \hat{a}_\textbf{k}^\dagger{u}
^{*}_k(\tau)e^{-i\textbf{k}\cdot\textbf{x}}\Big),
\end{equation}
and we find that the mode functions $u_k(\tau)$ satisfy the general equation of motion \cite{Muk12}
\begin{equation} \label{Muk20}
u''_k(\tau)+\omega^2_k(\nu, \tau){u_k(\tau)}=0,
\end{equation}
where $\omega_k(\nu, \tau)$ is frequency term that is dependent on the both of time and Hankel function index $\nu$ for general curved space-time, but is only time dependent for special dS space-time. Each solution $ u_k(\tau)$ must be normalized for all times according to
 \begin{equation}
  \label{nor11}
u^{*}_{k}u^{'}_{k}-u^{*'}_{k}u_{k}=-2i.
 \end{equation}
 A standard Bogoliubov coefficients $\beta_k$ \cite{Muk12, Gri13, Mam14}, can be related to the final phrase for the number of created particles $\langle N \rangle$ in the $ k $ mode as
\begin{equation} \label{Num1}
\langle N \rangle=|\beta_k|^{2}=-\frac{1}{2}+\frac{1}{4\omega_{k}(\nu, \tau)}|u'_{k}|^{2}+\frac{\omega_{k}(\nu, \tau)}{4}|u_{k}|^{2}.
 \end{equation}
 The concentration of created particles is readily obtained by integrating over all the modes.
\section{General Asymptotic-dS Modes Instead Special dS Mode}
In a pure dS background with $\nu=1.50$, the exact solution of (\ref{Muk20}) becomes
\begin{equation} \label{equ6}
u_{k}^{dS}=\frac{A_{k}}{\sqrt{k}}\Big(1-\frac{i}{k\eta}\Big)e^{-ik\eta}+\frac{B_{k}}{\sqrt{k}}\Big(1+\frac{i}{k\eta}\Big)e^{+ik\eta} .
\end{equation}
Where $A_{k}$ and $B_{k}$  are Bogoliubov coefficients. This solution by setting $A_{k}=1$  and $B_{k}=0$, leads to the Bunch-Davies (BD) mode

\begin{equation} \label{equ7}
u_{k}^{BD}=\frac{1}{\sqrt{k}}\Big(1-\frac{i}{k\eta}\Big)e^{-ik\eta}.
\end{equation}
 Since,the recent CMB results from Planck satellite impose an interesting constraint on the value of scalar spectral index approximately to be  $n_s = {0.968} \pm {0.006}$ at 95\% CL \cite{Obs6}.  This constraint indicates that the index of Hankel function $\nu$ lies in the range of $1.513\leq \nu \leq 1.519$ and at the early universe, space-time can indeed be replaced by a quasi-dS or \emph{nearly dS space-time} rather than dS space-time. So that any primitive excited or \emph{approximate dS space-time} can be considered as an acceptable state for initial background. Motivated by this fact that the inflation starts in approximate-dS Sitter space-time, we change the equation (\ref{Muk20}) to the general form \cite{man1},
\begin{equation} \label{Muk25}  \upsilon''_{k}+(k^{2}-\frac{2A}{\tau^2})\upsilon_{k}=0, \end{equation}
where $A$ is given by \cite{man1},
 \begin{equation}
 \label{alf26}  A=A(\nu)=\frac{4\nu^2-{1}}{8}.
 \end{equation}
  The general solutions of mode equation (\ref{Muk25}) can be written as \cite{Lin9, man1, Bau17}: \begin{equation}
\label{Han22}
\upsilon_{k}=\frac{\sqrt{\pi \tau}}{2}\Big(A_{k}H_{\nu}^{(1)}(|k\tau|)+B_{k}H_{\nu}^{(2)}(|k\tau|)\Big), \end{equation} where
$ H_{\nu}^{(1, 2)} $ are the Hankel functions of the first and second kind, respectively.
Let us consider the general form of the mode function by using the asymptotic expansion of Hankel function up to the higher order of ${1}/{|k\tau|}$ for the far past time limit as follow,
 \begin{equation} \label{gen27}  \upsilon^{gen}_{k}(\tau, \nu)=A_{k}\frac{e^{-{i}k\tau}}{\sqrt{k}}\big(1-i\frac{A}{k\tau}-\frac{\beta}{k^2\tau^2}-...\big)+
 B_{k}\frac{e^{{i}k\tau}}{\sqrt{k}}\big(1+i\frac{A}{k\tau}-\frac{\beta}{k^2\tau^2}+...\big),
 \end{equation}
note that $\beta={A(A-1)}/{2}$. The positive frequency solutions of the mode equation (\ref{Muk25}) are given by \cite{man1, Han11}
\begin{equation}
\label{mod28}  \upsilon_{k}=\frac{e^{-{i}k\tau}}{\sqrt{k}}\left(1-i\frac{A}{k\tau}-\frac{\beta}{k^2\tau^2}-...\right).
 \end{equation}
Note that, this general solution, only for $\nu=1.50$ reduces to the pure dS mode which results in the first (linear) order in terms of $1/{k\tau}$ and for another values of index $\nu$ , the solutions contain higher order in terms of $1/{k\tau}$. As we know the recent observational results indicate that our universe starts in an approximate dS or quasi-dS space-time [8, 9] with $\nu\simeq1.50$ and varying Hubble parameter.

\section{Calculation with Asymptotic-dS Background}
\subsection{Constraint on Coupling Constant from Scalar Spectral Index}
For the minimal coupled massless inflaton field in the pure dS background, we can write $\omega_{k}$ as,
\begin{equation}
\omega^2_k(\tau)={k}^2-\frac{2}{\tau^2}.
\end{equation}
On the other hand, for the minimal massless field in the quasi-dS dynamical background with $\nu\neq{1.50}$ , let us write $\omega^2_k(\tau)$ as follows \cite{man1, Bun8, Han11},
\begin{equation}
\label{tdp}
\omega^2_k(\tau)={k}^2-\frac{2A}{\tau^2}.,
\end{equation}

Only for the case of the pure dS space-time with $\nu=1.50$, we have $A=1$ and for the quasi-dS space-time for example quasi-dS space-time with  $\nu\approx1.50 +\epsilon$, where $\epsilon$ is slow-roll parameter, we have $\nu\neq1.50\quad or\quad A\neq1$.\\
Also from (\ref{equ82}), we offer the following frequency equation for non-minimal massless inflaton field in the pure dS background\cite{Per15},
\begin{equation}
\label{mdp}
\omega^2_k(\tau)={k}^2+(6\xi-1)\frac{2}{\tau^2}.
\end{equation}
If we compare (\ref{tdp})and (\ref{mdp}), we can consider the $\nu$-dependent term $\frac{2A}{\tau^2}$ with minimal coupled field $\xi=0$ in the quasi-dS background, as the $\xi$-dependent term ${\frac{2(1-6\xi)}{\tau^2}}$ with non-minimal coupling field $\xi\neq0$ in the dS background. So, by imposing the constraint on the value of scalar spectral index from \cite{Obs6}, we obtain the following constraint for spectral index, Hankel index and coupling constant in quasi-dS inflation,
\b
\label{Spec} 0.962 \leq n_s \leq 0.974,\e
\b 1.513 \leq  \nu \leq 1.519,\e
and
\b {-0.0047} \leq  \xi \leq {-0.0033}.\e
This finding is similar to the results that previously obtained in Ref.\cite{Tsu11} and in Ref. \cite{Han12} via different methods\cite{Far19}.\\
   Consequently, the field $u$ in the quasi-dS space-time obeys the same equation of motion for a non-minimal scalar field in the dS space-time that originate from the change of $\nu$, e.g. background space-time. Non-minimal coupling of scalar field with gravitational field is necessary in many situations of physical and cosmological interest. In most theories used to describe inflation, a non-vanishing value of the coupling constant is unavoidable\cite{Far18, Far19}. Therefore, this $\nu$-dependent term interpreted as the interaction between the scalar field and the gravitational field. As a result, the energy of the field $u$ is not conserved in this background, and more important, its quantization leads to particle creation at the expense of the time-dependent gravitational background\cite{Bir2, Per15}.
\subsection{Power Spectrum with Higher order Corrections in Quasi-dS space-time}
We use the following two-point function to calculate the power spectrum, \cite{man1, Bau17},
\begin{equation}
 \label{pow16} \langle0|\hat{u}_{k}(\tau)\hat{u}_{k'}(\tau)|0\rangle=\frac{1}{2}(2\pi)^{3})\delta^{3}(k+k')|u_{k}(\tau)|^2,
\end{equation}
where $\hat{u}$ is the quantum mode function. If we use the curvature perturbation as ${\cal R}_{k}(\tau)=\frac{u_{k}(\tau)}{a}(\frac{H}{{\dot{\bar{\phi}}}})$, we can calculate power spectrum $P_{{\cal R}}$ and the dimensionless power spectrum $\Delta_{{\cal R}}^{2}$ in terms of ${\cal R}_{k}(\tau)$ as follows\cite{Bau17},
\begin{equation}
 \label{pow17} \langle\hat{{\cal R}}_{k}(\tau)\hat{{\cal R}}_{k'}(\tau)\rangle=(2\pi)^{3}\delta^{3}(k+k')P_{{\cal R}},
\end{equation}
\begin{equation}
 \label{pow18} \Delta_{{\cal R}}^{2}=\frac{k^3}{2\pi^{2}}P_{{\cal R}}.
 \end{equation}
By using equations (\ref{pow17} -\ref{pow18}), for general modes (\ref{mod28}) up to second order of $1/k\tau$, the modified power spectrum for initial fixed time $\tau_0$\footnote{The time $\tau_0 =\frac{M_{Pl}}{kH}$ is a fixed initial time that U. H. Danielsson, first, used this fixed time for calculation of power spectrum \cite{Dan19}. For a given $k$, he choose a finite $\tau_0$ such that the physical momentum corresponding to $k$ is given by some fixed scale $M_{Pl}$. $M_{Pl}$ is the energy scale of new physics, e.g. the Planck scale.}, obtained in the following form\cite{man1},
  \begin{equation} \label{del36}
 \Delta_{{\cal R}}^{2}=\frac{H^2}{(2\pi)^{2}}(\frac{H^2}{\dot{\bar{\phi}}^2})\big[\frac{2\nu+1}{2(2\nu-1)}+(2\nu+1)^{2} \frac{(4\nu^{2}-9)}{64}\frac{H^{2}}{{M_{Pl}}^{2}}+...\big].
\end{equation}
 If we consider the vacuum in the infinite past time $\tau\rightarrow{0}$, the vacuum is exact, but if it is imposed at a later time, for example $\tau_0$, it is natural to expect non-vanishing magnitude of the correction term $(\sim \frac{1}{k\tau_0}=\frac{H}{M_{Pl}})$. Also, for pure dS space-time, e.g. $\nu=1.50$, this correction term is vanish and we will obtain scale-invariant power spectrum, but the corrections term is very tiny during inflation with quasi-dS space-time with $\nu=1.50+\epsilon$.

\subsection{Particle Creation in Quasi-dS Space-time}
In general, one expects particles to be created due to the changing gravitational field in an expanding universe. As a matter of fact the amount of created particles depends on how one actually fixes the vacuum \cite{Bir2, Ful3}. Hence at the first special time, we have a special observer which can assume the vacuum state that defined by pure dS mode. As a special observer $O_{one}$, we call such a vacuum the "one"-vacuum, $\hat{a}_{one}|one\rangle = 0$, \emph{because, in this time and other later time we have different points of space-time and in any point of space-time, there is a special observer frame and each observer in any frame may define for him/herself a special vacuum.} . At an arbitrary another moment of time, an observer would use the solution (\ref{mod28}) (with the $\nu\neq 1.50$) to define another vacuum, that we call "two"-vacuum $|{two}\rangle$. Because the quasi-dS mode depends on $\nu$ and such modes change with time, any selection of $\nu$ yields a different set of space-time and the different vacuum and hence, a set of different annihilation operators $\hat{a}_{two}$. So, we have, $\hat{a}_{two}|two\rangle= 0$. It is obvious that the "One" and "Two" vacuum built over the dS and the quasi-dS modes (or vice versa), are different. In particular, the "one"-vacuum $|{one}\rangle$ contains particles of the "two"-vacuum $|{two}\rangle$. So, by using equation (\ref{Num1}), we can calculate the number of created particles with the general asymptotic-dS modes as the following general form,
$$ \langle N(\nu,\tau)\rangle_{gen}=-\frac{1}{2}+\frac{1}{4}|k^{2}\tau^{2}-2A|^{1/2}\big[\frac{1}{k\tau}+\frac{A}{k^{3}\tau^{3}}
+\frac{B^{2}}{k^{5}\tau^{5}}+...\big]$$
\begin{equation}
\label{Num2}
+\frac{1}{4|k^{2}\tau^{2}-2A|^{1/2}}\big[k\tau-\frac{A}{k\tau}+\frac{(A-B)^{2}}{k^{3}\tau^{3}}
+\frac{4B^{2}}{k^{5}\tau^{5}}+...\big].
\end{equation}
As a important result, we see that the number of created particles relate to index $\nu$, that can be means the creation of particles may be dependent to the observer, background space-time and it's dynamics. Also, it can grows in the later time limit $\tau\rightarrow{0}$, which particle creation appears to be significant, that has been obtained in \cite{Per15}. But these correction terms is very tiny for early time limit and become zero in the infinite past, $\tau\rightarrow{-\infty}$.

\section{Discussions in Some Specific Background}
To verify the general formula (\ref{Num2}), let us study particle creation at initial fixed time $\tau_0$, for some specific backgrounds such as the Minkowski, dS and the quasi-dS space-times in early universe.
\subsection{Minkowski Background with Conformal Coupling}
If we consider $\nu=0.50$, we obtain $A=0$, $\xi=1/6$ and $B=0$, so the general mode (\ref{mod28})reduces to the Minkowski space-time mode,
\begin{equation} \label{mod281}  u^{Min}_{k}=\frac{e^{-{i}k\tau}}{\sqrt{k}},
 \end{equation}
and it is clear that there is no particle creation which is the standard result \cite{Per15},
\begin{equation} \langle{N(0.50,\tau)}\rangle_{Min}=0.
\end{equation}
\subsection{dS Background with Minimal Coupling}
If we consider $\nu=1.50$, then we obtain $A=1$, $\xi=0$ and $B=0$, so the general mode (\ref{mod28}) reduces to the dS space mode with BD mode (\ref{equ7}) and one obtains
 $$ \langle{N(1.50,\tau)}\rangle_{BD}=-\frac{1}{2}+\frac{1}{4}|k^{2}\tau^{2}-2|^{1/2}(\frac{1}{k\tau}+\frac{1}{k^{3}\tau^{3}})$$
 \begin{equation}+\frac{1}{4|k^{2}\tau^{2}
-2|^{1/2}}(k\tau-\frac{1}{k\tau}+\frac{1}{k^{3}\tau^{3}}).
\end{equation}
This exhibits that in this background massless particles can be produced, the similar result previously was argued in Ref.\cite{Per15}.

\subsection{Quasi-dS Background with Non-minimal Coupling}
 In this final subsection \emph{we show that the massless minimal coupled particles in the quasi-dS background can be equivalent to the massless non-minimal coupled or massive minimal coupled in the dS background}.\\
 The case of $\nu\neq 1.50$ is interesting since it can be related to the mass-dependent parameter in the context of f(R)-gravity through  $\nu=\sqrt{9/4-12\alpha{m^2}}$, where $\alpha$ is a dimension-full parameter which appears as a coefficient in f(R)-gravity \cite{Per15}. On the other hand this case, as mentioned, can be related to the quasi-dS modes in which $\nu=\sqrt{9/4-12{\xi}}$, appears as a variable parameter extracting by the changing of the background space-time. Therefore, within the context of the particle creation one can deal with massive particles creation. For example we study the quasi-dS space-time, that observationally is very important and in that case we have $\nu=1.50+\epsilon$, and by considering $\epsilon=0.013$, we obtain $A\simeq1.015$ , $B\simeq0.008$, and the number of the created massive particles, up to second order $\frac{1}{k\tau}$ of quasi-dS modes, is given by
 $$ \langle N(1.513,\tau)\rangle_{gen}=$$
 $$ -\frac{1}{2}+\frac{1}{4}|k^{2}\tau^{2}-2.03|^{1/2}\big[\frac{1}{k\tau}+\frac{1.015}{k^{3}\tau^{3}}
+\frac{0.00006}{k^{5}\tau^{5}}+...\big]$$
\begin{equation} +\frac{1}{4|k^{2}\tau^{2}-2.03|^{1/2}}\big[k\tau-\frac{1.015}{k\tau}+\frac{1.014}{k^{3}\tau^{3}}+\frac{0.0003}{k^{5}\tau^{5}}+...\big].
\end{equation}
Note that the correction terms obtained from the above results grow in the later time but it is very tiny at the far past time limit, specially at initial fixed time $\tau_0$. In the infinite far past, all of  the correction terms are zero.
\section{Conclusions}
Actually, Recent Planck2015 results motivated us to use non-Bunch-Davies vacuum. In this paper, we used the general form of quasi-dS modes as non-trivial initial states during inflation to calculate the expectation value of the created particles in terms of the Hankel function index and the conformal time. First, we consider the mode in pure dS space-time as a background state and for the excited states we proposed general quasi-dS modes. For general form of quasi-dS modes, e.g. $\nu\neq 1.50$, we use the asymptotic expansion of the Hankel function to the higher order of approximation. By taking into account this alternative modes and the effects of trans-Planckian physics, we obtained the expectation value of the created particles in standard approach. For the quasi-dS the higher order corrections of it is non-linear, but it converge at early time, and in minkowski and dS limit corrections reduce to the form that obtained from several previous conventional methods.\\
On the other hand, we compared the general form of inflaton field equation (\ref{tdp}) and (\ref{mdp}), and the Planck 2015 constraint on spectral index , and we obtained the possible new constraints (\ref{Spec}), for the values of coupling constant in the era with quasi-dS space-time. This finding is similar to the results that previously obtained by other scientists via different methods. As an main result, we indicated that the number of created particles and the value of coupling constant can be dependent to the selection of background space-time and it's dynamics.\\
Consequently, by slightly deviation from the pure dS mode to the quasi-dS modes, we must have $\nu\neq 1.50$, and this value stimulated us to using asymptotic-dS modes with higher order term of $\frac{1}{k\tau}$, which lead to the converged higher order trans-Planckian corrections term at initial fixed time $\tau_0$ and large number of non-minimal particles at later time. As it is known at the early time in short distance regime the energy was very high and the universe was in the maximal symmetry, therefore these tiny higher order trans-Planckian corrections and created particles with non-minimal coupling constant may be recast as a source to generate an effective interaction between quantum fields and gravity. In the context of effective field theory, in micro scales the initial symmetry may be broken by such non-linear corrections to outburst and propagation non-minimal quantum field in the macro scale, and in later time the large number of created non-minimal particles can be to interact with gravity in macro scales to formation of galaxies and galaxies cluster in large scale at later time. In future work, the consequences of the constraint on the coupling constant and other result from quasi-dS space-time with dynamical background will be investigated.

\noindent {\bf{Acknowlegements}}: We would like to thank S. A. Fulling for useful comment in the subject of non-minimal coupling and M. V. Takook, K. nozari and H. Pejhan for useful and serious discussions. This work has been supported by the Islamic Azad University, Ayatollah Amoli Science and Research Branch, Amol,Mazandaran, Iran.

\end{document}